\documentclass[article,twocolumn]{revtex4}
\usepackage{amsfonts,amsmath,amssymb,latexsym,epsfig}

\usepackage{color,hyperref}
\hypersetup{
    colorlinks,%
    citecolor=blue,%
    filecolor=blue,%
    linkcolor=blue,%
    urlcolor=blue
}

\newcommand{\pr}[1]{\left( #1\right)}
\newcommand{\prr}[1]{\left[ #1 \right]}
\newcommand{\es}[1]{\begin{equation}\begin{split}#1\end{split}\end{equation}}

\newcommand{\R}{\mathbb{R}}

\newcommand{\f}{\mathrm{free}}
\newcommand{\bsn}{\mathrm{b}}
\newcommand{\frm}{\mathrm{f}}
\newcommand{\ie}{\textit{i.e.} }

\begin{document}
\title{Influence of boundary conditions on quantum escape}
\author{Orestis Georgiou$^{1}$, Goran Gligori\'{c}$^{1,2}$, Achilleas Lazarides$^{1}$, Diego F.M. Oliveira$^{3}$, Joshua D. Bodyfelt$^{1}$, Arseni Goussev$^{1}$}
\affiliation{1 Max Planck Institute for the Physics of Complex Systems - N\"{o}thnitzer Stra{\ss}e 38, 01187, Dresden, Germany.}
\affiliation{2 Vin\v{c}a Institute of Nuclear Sciences, University of Belgrade - P.O. Box 522, 11001 Belgrade, Serbia.}
\affiliation{3 Institute for Multiscale Simulations - Friedrich-Alexander Universit\"{a}t - Naegelsbachstrasse 49b, D-91052, Erlangen, Germany.}
\begin{abstract}
It has recently been established that quantum statistics can play a crucial role in quantum escape.
Here we demonstrate that boundary conditions can be equally important -- moreover, in certain cases, may lead to a complete suppression of the escape.
Our results are exact and hold for arbitrarily many particles.
\end{abstract}

\maketitle

\section{Introduction and setup}
\label{sec:setup}

The question of how particles escape from a partially confining region has long been at the center of many experimental and theoretical studies, leading to a number of profound
discoveries in mathematical physics (for a recent review see Ref. \cite{APT12}). Most of the theoretical progress has been made in the context of classical mechanics, where different dynamical behaviors --
regular, chaotic, or mixed -- lead to different escape laws; typically quantified by the decay of the survival (\ie non-escape) probability. On the quantum mechanical side, there
has been a surge of renewed interest in the escape properties of few-particle systems \cite{TS11,TS11b,AC11,GL11,HF12,WGGR08}, in connection with recent advances in experimental
control and manipulation of a small number of ultra-cold atoms \cite{SZLOWJ11,ZSLWRBJ12}. In particular, particle-particle interactions and quantum statistics have been shown to
significantly influence the escape. Here, we extend the theory to account for the influence of boundary conditions.

\begin{figure}[h]
\begin{center}
\includegraphics[scale=0.63]{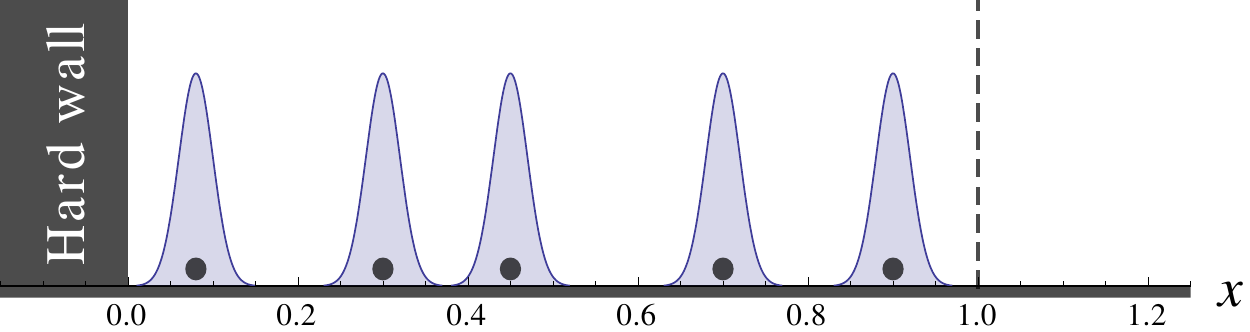}
\caption{\label{fig:ic} Schematic plot of the initial configuration for $N=5$ particles represented as Gaussian wave packets.
}
\end{center}
\end{figure}

We consider a system of $N$ non-interacting particles in one spatial dimension which are initially ($t=0$) confined within the interval $I = (0,1)$ with a hard impenetrable wall
placed at the origin. A schematic plot of the setup is shown in Fig.\ref{fig:ic} for $N=5$ particles represented as Gaussian wave packets.
%%%%%%%%%%%%%%%%%%%%%%%%%%%%%%%%%%%%%%%%%%%%%%%%%%%%%%%%%%%%%%%%%%%%%%%%%%%%%%%%%%%%%%%%%%%%%%%%%%%%%%%%%%%
%%%%%%%%%%%%%%%%%%%%%%%%%%%%%%%%%%%%%%%%%%%%%%%%%%%%%%%%%%%%%%%%%%%%%%%%%%%%%%%%%%%%%%%%%%%%%%%%%%%%%%%%%%%
For positive times, $t > 0$, the particles are allowed to explore the positive real line. The quantum state of the system is described by the $N$ particle wave function $\Psi(x_1,
\ldots, x_N, t)$, which satisfies the time-dependent Schr\"{o}dinger equation,
\begin{equation}
i\frac{\partial}{\partial t} \Psi(x_1, \ldots, x_N, t) = \hat{H}^{(N)} \Psi(x_1, \ldots, x_N, t) \label{SE}
\end{equation}
where $\hat{H}^{(N)}= - \frac{1}{2} \left( \frac{\partial^2}{\partial x_1^2} + \ldots + \frac{\partial^2}{\partial x_N^2} \right)$ is the $N$-particle Hamiltonian operator and we have set the Planck's constant and particle masses to unity, i.e., $\hbar = m_1 = \ldots m_N = 1$.
The initial wave function is such that $\Psi(x_1, \ldots, x_N, t=0) = 0$ if $x_j\not\in I$ for all $j\in[1,N]$.

Under time evolution, the wave function diffuses outside $I$. One way to measure this is through the survival probability $P^{(N)}(t)$, \ie the probability that all $N$ quantum
particles are in $I$ at times $t>0$. In terms of the wave function, the survival probability is defined as
\begin{equation}
P^{(N)}(t)= \int_{ I^{N}} | \Psi(x_1, \ldots, x_N, t) |^2 d x_1 \ldots d x_N . \label{survival}
\end{equation}

Quantum statistics obeyed by the $N$ particles -- whether bosonic or fermionic -- can have a dramatic effect on the time-decay of the survival probability. More specifically, it
was observed in Ref.~\cite{TS11} that the survival probability $P^{(2)}(t)$ asymptotically decays to zero as $\sim t^{-6}$ for bosons and $\sim t^{-10}$ for fermions.
This surprisingly simple, yet generic, result was obtained by an asymptotic expansion of the appropriate quantum propagator for large $t$. This observation was later generalized in
Ref.~\cite{AC11} for $N$ particles, showing that $P^{(N)}(t)$ decays like $\sim t^{-3N}$ and $\sim t^{-N(2N+1)}$ for bosons and fermions respectively. The faster fermionic decay
was attributed to the effective hard-core interaction among fermions causing anti-bunching and hindering the reconstruction of the initial state. In both works \cite{TS11,AC11}
Dirichlet boundary conditions were imposed at the hard wall, \ie $\Psi(x_1, \ldots, x_N, t) \big|_{x_j = 0} = 0$ for all $j\in[1,N]$.

When a particle is constrained to move within a limited volume, Dirichlet ($\mathcal{D}$) or Neumann ($\mathcal{N}$) boundary conditions (BCs) are usually employed in order to reflect local probability
preservation at $x=0$. It is well known however that the general solution to this requirement is described by a Robin ($\mathcal{R}$) BC \cite{GA98}, defined as a weighted combination of the $\mathcal{D}$ and $\mathcal{N}$ BCs
\begin{equation}
\left( \frac{\partial}{\partial x_j} - \eta \right) \Psi(x_1, \ldots, x_N, t) \Big|_{x_j=0} = 0 \,, \quad \forall j, \label{BC}
\end{equation}
and controlled by the parameter $\eta\in\R$, such that $\eta=0$ corresponds to the $\mathcal{N}$ and $\eta=\infty$ to the $\mathcal{D}$ BCs.
The parameter $\eta$ has the physical interpretation of a phase shift of the wave function on reflection with the wall at the origin.
Therefore, the following natural question arises: \textit{What is the
influence of boundary conditions on the decay of the survival probability?} More specifically, {\it how does the survival probability $P^{(N)}(t)$ depend on $\eta$?} This we
address in the current setting by deriving and analyzing the exact quantum propagator.
It is worth noting that $\mathcal{R}$ BCs are commonly used in Sturm-Liouville type problems which appear in many contexts in engineering and applied mathematics including for example in acoustics and in convection-diffusion processes.

We motivate this investigation by considering the simplest case of a single particle, $N=1$. The dynamical evolution of the wave function can be described by the quantum propagator
$K_{\eta}(x,x',t)$ as
\begin{equation}
\Psi(x, t)= \int_I K_{\eta}(x,x',t) \Psi(x', 0) \, d x' . \label{prop}
\end{equation}
The propagator \cite{B89} is a generalized function that satisfies the time-dependent Schr\"{o}dinger equation
\begin{equation}
\pr{i \frac{\partial}{\partial t} + \frac{1}{2} \frac{\partial^2}{\partial x^2} } K_{\eta}(x,x',t) = 0, \label{KE}
\end{equation}
the initial condition
\begin{equation}
 \lim_{t \rightarrow 0}  K_{\eta}(x,x',t) = \delta(x-x') \,, \label{propagator-initial-condition}
\end{equation}
the $\mathcal{R}$ BC at $x=0$
\begin{equation}
\left( \frac{\partial}{\partial x} - \eta \right) K_{\eta}(x,x',t) \Big|_{x=0} = 0 \,, \label{propagator-boundary-condition}
\end{equation}
and that vanishes as $x \rightarrow \infty$ at negative imaginary times, i.e., $\lim\limits_{x \rightarrow \infty} K_{\eta}(x,x',-i T) = 0$ for $T > 0$.

The cases of $\eta=0$ ($\mathcal{N}$ BC) and $\eta=\infty$ ($\mathcal{D}$ BC) are particularly simple. Using the method of images we can immediately write down the full single-particle propagator as
\begin{eqnarray}
K_{0}(x,x',t) &=& K_{\f}(x,x',t) + K_{\f}(x,-x',t) \,, \label{images_N} \\
K_{\infty}(x,x',t) &=& K_{\f}(x,x',t) - K_{\f}(x,-x',t) \,, \label{images_D}
\end{eqnarray}
where
\begin{equation}
K_{\f}(x,x',t) = \frac{1}{\sqrt{2\pi i t}} \exp \left(-\frac{(x-x')^2}{2 i t} \right) \label{free_propagator}
\end{equation}
is the free-particle propagator, \ie the propagator describing the motion on the real line without a reflecting wall at the origin.

The asymptotic decay of $P^{(1)} (t)$ is entirely determined by the long-time expansion of the quantum propagator. Thus,
\begin{equation}
K_0(x,x',t)= \sqrt{\frac{2}{i \pi t}} + \mathcal{O}(t^{-3/2}) \label{KN}
\end{equation}
and
\begin{equation}
 K_{\infty}(x,x',t)=-\sqrt{\frac{2 i}{\pi}}\frac{x x'}{t^{3/2}} + \mathcal{O}(t^{-5/2}) \,, \label{KD}
\end{equation}
entailing
\es{
  \mathrm{\mathcal{N} \; BC:} \; \; &P^{(1)} \sim t^{-1} \,,  \\
  \mathrm{\mathcal{D} \; BC:} \; \; &P^{(1)} \sim t^{-3} \,. \label{NDasy}
}
So for the single-particle case, the $\mathcal{D}$ BC leads to faster escape.

When addressing the case of two particles, $N=2$, the issue of quantum statistics comes into play. For noninteracting particles, there are two ways of taking into account bosonic
(b) or fermionic (f) statistics: either (i) by appropriately symmetrizing the initial state of the system,
\begin{equation}
\Psi^{(\bsn / \frm)}(x_1,x_2,0) = \frac{1}{\sqrt{2}} \prr{ \psi_1(x_1) \psi_2(x_2) \pm \psi_1(x_2) \psi_2(x_1) }, \label{psisym}
\end{equation}
with
``$+$'' corresponding to bosons and ``$-$'' to fermions, and propagating it with $K_{\eta}(x_1,x'_1,t) K_{\eta}(x_2,x'_2,t)$, or (ii) by introducing an effective, appropriately
symmetrized two-particle propagator
\begin{align}
  K^{(\bsn / \frm)}_{\eta} (x_1,x_2,&x'_1,x'_2,t) \nonumber\\ =
  \frac{1}{\sqrt{2}} \Big[ &K_{\eta}(x_1,x'_1,t) K_{\eta}(x_2,x'_2,t)
  \nonumber\\ &\pm K_{\eta}(x_2,x'_1,t) K_{\eta}(x_1,x'_2,t) \Big] \,,
\label{K2}
\end{align}
and applying it upon the product state $\psi_1(x_1) \psi_2(x_2)$. Indeed, the two-particle state at $t>0$ can be written as
\begin{align}
  &\Psi^{(\bsn / \frm)} (x_1, x_2, t) \nonumber \\ &= \int_{I^2}
  K_{\eta} (x_1,x'_1,t) K_{\eta} (x_2,x'_2,t) \, \Psi^{(\bsn /
    \frm)}(x'_1,x'_2,0) \, d x'_1 d x'_2 \nonumber \\ &= \int_{I^2}
  K^{(\bsn / \frm)}_{\eta} (x_1,x_2,x'_1,x'_2,t) \, \psi_1(x'_1)
  \psi_2(x'_2) \, d x'_1 d x'_2 \,,
\label{Psi2}
\end{align}
where the first equality corresponds to (i) and the second to (ii). Being interested in the asymptotics of the survival probability (and inspired by Ref.\cite{TS11}),
we adopt approach (ii), as it allows us to focus only on the kernel, as opposed to the full integral. In other words, the long-time decay of $P^{(2)}(t)$ is determined by $
K^{(\bsn / \frm)}_{\eta} (x_1,x_2,x'_1,x'_2,t)$ in the limit $t \rightarrow \infty$. For $\eta = 0$ and $\eta = \infty$ one obtains
\es{
\mathrm{Bosons, \; \mathcal{N} \; BC:} \;\; &K_0^{(\bsn)} \sim t^{-1} \;\; \Rightarrow \; P^{(2)} \sim t^{-2} \,, \\ \mathrm{Fermions, \; \mathcal{N} \; BC:} \;\; &K_0^{(\frm)} \sim t^{-3} \;\;
\Rightarrow \; P^{(2)} \sim t^{-6} \,,  \\ \mathrm{Bosons, \; \mathcal{D} \; BC:} \;\; &K_\infty^{(\bsn)} \sim t^{-3} \;\; \Rightarrow \; P^{(2)} \sim t^{-6} \,,  \\ \mathrm{Fermions, \; \mathcal{D}
\; BC:} \;\; &K_\infty^{(\frm)} \sim t^{-5} \;\; \Rightarrow \; P^{(2)} \sim  t^{-10} \,.
\label{eq18}}

It is interesting to note that the asymptotic decay of the survival probability of two fermions with the $\mathcal{N}$ BC is governed by the same exponent as that of two bosons with the $\mathcal{D}$ BC.
This simple observation clearly illustrates that BCs imposed at the perfectly reflecting wall at the origin are as important for the particle escape as quantum statistics. It also
suggests that by tuning the value of $\eta$ one may expect to probe the continuum of possible decay rates, ranging between those for the $\mathcal{D}$ and $\mathcal{N}$ BCs. However, as we shall show in
the following sections, this is not the case, as the decay exponent turns out to be a discontinuous function of $\eta$.

\section{Single-particle case}
\label{sec:single}

Unlike in the special cases of $\eta=0$ ($\mathcal{N}$ BC) and $\eta = \infty$ ($\mathcal{D}$ BC), the quantum propagator for an arbitrary $\eta$ can not be straightforwardly obtained using the method of
images; alternative techniques have to be employed. We begin our study of the single-particle survival probability by deriving an exact closed-form expression for $K_\eta
(x,x',t)$, defined through Eqs.~(\ref{KE}-\ref{propagator-boundary-condition}).

\subsection{Exact propagator}
One way to construct the single-particle propagator $K_\eta (x,x',t)$ is by using a complete orthonormal set of eigenstates $\{ \phi_k \}$ of the Hamiltonian $\hat{H}^{(1)}$,
\begin{equation}
 -\frac{1}{2} \frac{d^2}{d x^2} \phi_k(x) = E_k \phi_k(x) \,, \label{phi_definition}
\end{equation}
satisfying the $\mathcal{R}$ BC at the origin, $\left( \frac{d}{dx} - \eta \right) \phi_k(0) = 0$. As it
will become clear from the following discussion (and as originally noted in Ref.\cite{LM77}), there is an important spectral difference between the case of $\eta > 0$ and that
of $\eta < 0$. We analyze the two cases in succession.

In the case of $\eta > 0$, the energy spectrum of $\hat{H}^{(1)}$ is continuous. Indeed, the corresponding eigensystem is given by
\begin{equation}
 \phi_k(x)= \sqrt{\frac{2}{\displaystyle \pi \left( 1 + \frac{k^2}{\eta^2} \right)}} \left[ \sin(k x) + \frac{k}{\eta} \cos(k x) \right] \label{phi_k}
\end{equation}
with $E_k = k^2 / 2$ and $k > 0$. The Hamiltonian eigenstates for the $\mathcal{N}$ and $\mathcal{D}$ BCs are respectively obtained as the limits $\lim\limits_{\eta \rightarrow 0} \phi_k (x)
= \sqrt{\frac{2}{\pi}} \cos (k x)$ and $\lim\limits_{\eta \rightarrow \infty} \phi_k (x) = \sqrt{\frac{2}{\pi}} \sin (k x)$.

It can be verified by direct integration that the set of the eigenstates $\{ \phi_k \}$ is orthonormal,
\begin{equation}
\int_0^\infty \! \phi_k (x) \phi_{k'} (x) \, d x = \delta(k-k') \,, \label{phi_k-normalization}
\end{equation}
and complete,
\begin{equation}
\int_0^\infty \! \phi_k (x) \phi_k (x') \, d k = \delta(x-x') \,, \quad \eta > 0 \,. \label{phi_k-completeness}
\end{equation}
Consequently, the propagator, expressed in the basis formed by $\{ \phi_k \}$, reads
\begin{align}
  K_\eta (x,x',t) &= \langle x | e^{-i \hat{H}^{(1)} t} | x' \rangle
  \nonumber \\ &= \int_0^\infty \! \phi_k (x) \phi_k (x') e^{-\frac{1}{2} i k^2 t
 } \, d k \,, \quad \eta > 0 \,. \label{prop_in_phi_k_basis}
\end{align}
In view of Eq.~(\ref{phi_k}), the integral in Eq.~(\ref{prop_in_phi_k_basis}) can be rewritten as
\begin{align}
  &\frac{1}{2 \pi} \int_{-\infty}^{+\infty} \! \big[ \cos(x_+ k) +
    \cos(x_- k) \big] e^{-\frac{1}{2} i k^2 t } \, d k \nonumber \\ &\quad
  \quad -\frac{\eta}{\pi} \int_{-\infty}^{+\infty} \! \frac{\eta
    \cos(x_+ k) - k \sin(x_+ k)}{k^2 + \eta^2} \, e^{-\frac{1}{2} i k^2 t}
  \, d k \nonumber \\ & = \mathcal{I}_1 - \mathcal{I}_2 \nonumber
\end{align}
with $x_{\pm} = x \pm x'$. Then, it is easy to see that the first term is nothing but the quantum propagator for the $\mathcal{N}$ BC, i.e., $\mathcal{I}_1 = K_0 (x,x',t)$. The
corresponding evaluation of the integral in the second term yields $\mathcal{I}_2 = \eta \, \textrm{erfc} \left( \frac{x_+ + i \eta t}{\sqrt{2 i t}} \right) e^{ \eta x_+
+ \frac{1}{2} i t \eta^{2} }$, where ``$\textrm{erfc}$'' denotes the complementary error function. Combining the two terms we obtain our first main result
\begin{align}
  K_\eta (x,&x',t) = K_0 (x,x',t) \nonumber \\ &-\eta \, \textrm{erfc}
  \left[ \frac{x+x' + i \eta t}{\sqrt{2 i t}} \right] e^{ \eta (x+x')
    + \frac{1}{2}i t \eta^{2} } \,.
\label{Keta}
\end{align}

In the case of $\eta < 0$, the set of eigenstates $\{ \phi_k \}$ no longer forms a complete basis \cite{LM77}.
Indeed, in addition to the unbound eigenstates, forming a continuous part of the energy
spectrum, there exists an isolated bound state
\begin{equation}
\chi(x) = \sqrt{2|\eta|} \, e^{-|\eta| x} \,, \quad \eta < 0 \,, \label{chi}
\end{equation}
corresponding to the negative energy $-\eta^2 / 2$, \ie $\hat{H}^{(1)} |\chi\rangle = -\frac{\eta^2}{2} |\chi\rangle$.
The state is normalized to unity, $\langle \chi | \chi \rangle = 1$, and is orthogonal to the
eigenstates of the continuous part of the energy spectrum, i.e.,
$\langle \chi | \phi_k \rangle = 0$ for any $k$.
The wave function
$\chi(x)$ is localized in the vicinity of the hard wall, indicating that the BCs here correspond to an effective zero-range attractive force.
%%%%%%%%%%%%%%%%%%%%%%%%%%%%%%%%%%%%%%%%%%%%%%%%%%%%%%%%%%%%%%%%%%%%%%%%%%%%%%%%%%%%%%%%%%%%%%%%%%%%%%%%%%%%%%%%%%%%%%%%%%%%%%%%%%5
\begin{figure*}
\begin{center}
 \includegraphics[scale=0.215]{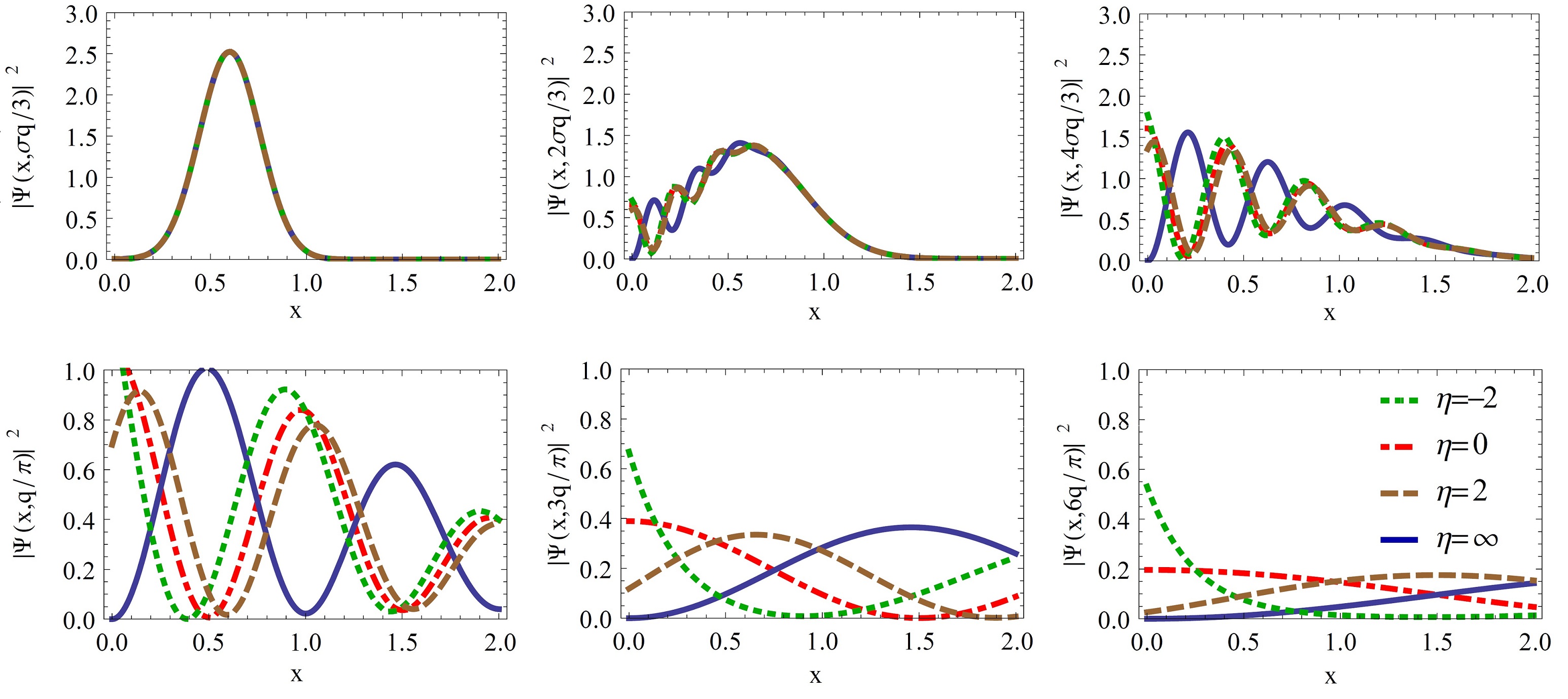}
\caption{\label{fig:wp6} (Color online) Snapshots of the evolution of the density $|\Psi(x,t)|^{2}$ of a Gaussian wave packet initially centered at $q=0.6$ with width $\sigma=0.1$
at times $t=t_w, 2 t_w, 4 t_w$, for the top three panels and at $t=t_a, 3 t_a, 6 t_a$, for the bottom three. The values $t_w=q\sigma/3$ and $t_a=q/\pi$ are respectively the
approximate times when the wave packets start interacting with the wall, and when the asymptotic regime sets in (see discussion in main text for details). The four curves in each
panel correspond to the evolution under different propagators $K_{\eta}$ with $\eta=\infty,0,2$, and $-2$ as indicated in the last panel.}
\end{center}
\end{figure*}
%%%%%%%%%%%%%%%%%%%%%%%%%%%%%%%%%%%%%%%%%%%%%%%%%%%%%%%%%%%%%%%%%%%%%%%%%%%%%%%%%%%%%%%%%%%%%%%%%%%%%%%%%%%%%%%%%%%%%%%%%%%%%%%%%%%%%%5

The presence of the bound state modifies the completeness relation
(cf. Eq.~(\ref{phi_k-completeness}))
\begin{equation}
 \int_0^\infty \! \phi_k (x) \phi_k (x') \, d k + \chi(x) \chi(x') = \delta(x-x') \,, \quad \eta < 0 \,, \label{phi_k-chi-completeness}
\end{equation}
so that the propagator expanded in terms of the complete basis $(\{ \phi_k \}, \chi)$ reads (cf. Eq.~(\ref{prop_in_phi_k_basis}))
\begin{align}
  K_\eta (x,x',t) = \int_0^\infty \! &\phi_k (x) \phi_k (x') e^{-\frac{1}{2} i k^2
    t} \, d k \nonumber\\ &+ \chi(x) \chi(x') e^{\frac{1}{2} i \eta^2 t }
  \,, \quad \eta < 0 \,. \label{prop_in_phi_k-chi_basis}
\end{align}
The evaluation of the last integral for $\eta < 0$ proceeds in the same manner as in the case of $\eta > 0$, and the resulting expression for the full propagator coincides with
that given by Eq.~(\ref{Keta}).
This concludes our derivation of the single-particle propagator.

At this point we remark that the problem of a single particle restricted to the positive real line by a hard wall, is equivalent to that of two hard-core particles on the full real line through a change of variables to center of mass and relative coordinates.
Moreover, the BCs imposed in the former correspond to the quantum statistics obeyed by the latter allowing for the possibility of anyons \cite{LM77}. The equivalence in the case of many particles is however more subtle with essential differences in higher dimensions and will not be discussed in the current paper.

\subsection{Short-time dynamics}
Knowledge of the exact propagator \eqref{Keta} allows us to study the full dynamical evolution of the single particle wave function $\Psi(x,t)$. In particular, in order to
understand the influence of BCs on the escape, it is instructive to apply $K_\eta$ to a spatially localized initial state given by a Gaussian wave packet
\begin{equation}
 \Psi(x,0)= \pr{\frac{1}{\pi \sigma^{2}}}^{\frac{1}{4}} e^{-\frac{(x-q)^{2}}{2 \sigma^{2}}} \label{gaussian}
\end{equation}
centered at $q\in(3\sigma,1-3\sigma)$ with spatial width $\sigma\ll1$, such that $\Psi(x\not\in I,0)\approx 0$.

Fig.\ref{fig:wp6} shows snapshots at different times of the probability density function $|\Psi(x,t)|^{2}$ under time evolution by $K_\eta$ for four different values of
$\eta$. We denote these four cases by: $(i)$ $\eta=\infty$, $(ii)$ $\eta=0$, $(iii)$ $\eta=2$ and $(iv)$ $\eta=-2$. For the first two simple cases $(i)$ and $(ii)$ (corresponding
to the blue and red curves in Fig.\ref{fig:wp6}), the density can be written down explicitly as
\es{
|\Psi(x,t)|^{2}&= \frac{2}{\sigma\sqrt{\pi}} \frac{e^{-\frac{\theta^{2}}{1+\tau^{2}}}}{\sqrt{1+\tau^2}}
e^{-\frac{\xi^{2}}{1+\tau^{2} }} \\
&\times \prr{\cosh\pr{ \frac{2  \xi \theta }{1+\tau^2} } \mp \cos\pr{ \frac{2\tau \xi \theta }{1+\tau^2} }},
\label{density}
}
where we have used the scaled variables $\tau=t/\sigma^{2}>0$, $\xi=x/\sigma$, and $\theta=q/\sigma$, for the sake of brevity. The ``$-$'' and ``$+$'' signs in \eqref{density}
correspond to cases $(i)$ and $(ii)$ respectively. Cases $(iii)$ and $(iv)$ in the figure were calculated through numerical integration of Eq.\eqref{prop}. Inspecting
Eq.\eqref{density} we can identify the first exponential as the main time envelope and its denominator as being indicative of the diffusive process. The cosine term describes the
oscillations due to interaction with the wall and are clearly seen in Figure \ref{fig:wp6}. Note that the amplitude of these oscillations is suppressed for small times.

We now qualitatively describe the dynamics observed in Fig.\ref{fig:wp6} with the aid of Eq.\eqref{density} and identify three distinctly different time scales. For times $0< t
\leq t_w$, all four curves coincide as they diffuse freely until the wave packets start interacting with the wall. This occurs when $3\sigma \sqrt{1-\tau^2}\approx q$, and so $t_w
\approx q\sigma/3$ for $\sigma\ll1$. For times $t> t_{w}$, oscillations appear due to reflection off the wall and interference occurs. The wall interaction however is different in
each case due to the different BCs prescribed by $\eta$. It is clear from the figure that $(i)$ and $(ii)$ are completely out of phase, while $(iii)$ and $(iv)$ closely follow
curve $(ii)$ and are only slightly out of phase. The oscillations spread until $t\approx t_a$, when there is only a single maximum of $|\Psi(x,t)|^{2}$ left in $I$ for
case $(i)$ and a single minimum for case $(ii)$. The time $t_a$ can be extracted from the cosine term in Eq.\eqref{density} by requiring its argument to be $2\pi$ at $x=1$. We thus
have that $t_a\approx q/\pi$ for $\sigma\ll1$. Indeed, at times greater than $t_a$, the asymptotic regime sets in and there are no more oscillations. Cases $(i)$ to $(iii)$ behave
essentially the same, as they diffuse at different rates -- while case $(iv)$ seems to ``stick'' to the wall rather than diffuse away. The difference between $(iii)$ and $(iv)$ can
be more clearly seen in the two GIF animation files included as supplementary material. We postpone further discussion on the long time behavior of the four cases for the next
subsection and turn to the survival probability function.
%%%%%%%%%%%%%%%%%%%%%%%%%%%%%%%%%%%%%%%%%%%%%%%%%%%%%%%%%%%%%%%%%%%%%%%%%%%%%%%%%%%%%%%%%%%%%%%%%%%%%%%%%%%%%%%%%%%%%%5
\begin{figure}[h!]
\begin{center}
\includegraphics[scale=0.58]{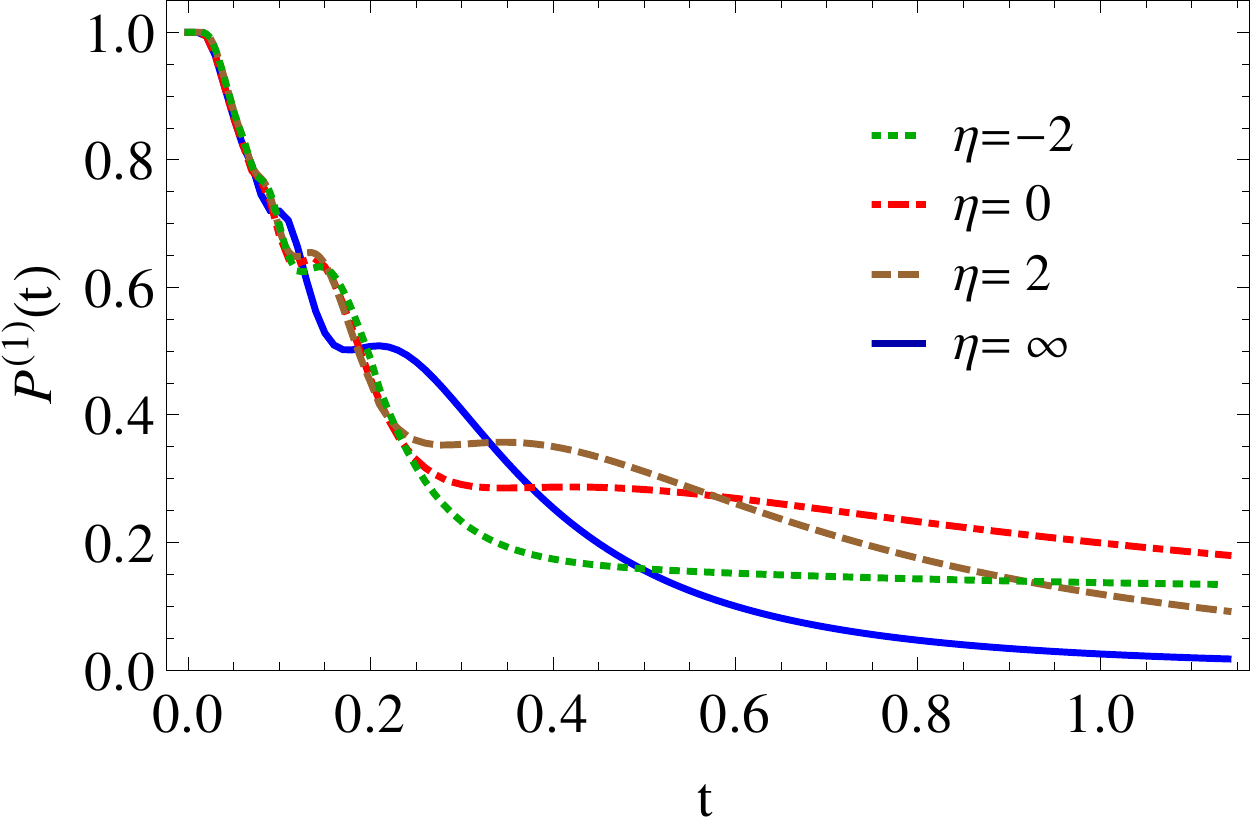}
\caption{\label{fig:pt} (Color online) The single particle survival probability in $I$ for the four cases described in Figure \ref{fig:wp6} using the same values for $q=0.6$ and
$\sigma=0.1$. The asymptotic regime sets in around $t\approx t_a=q/\pi\approx 0.191$.}
\end{center}
\end{figure}
%%%%%%%%%%%%%%%%%%%%%%%%%%%%%%%%%%%%%%%%%%%%%%%%%%%%%%%%%%%%%%%%%%%%%%%%%%%%%%%%%%%%%%%%%%%%%%%%%%%%%%%%%%%%%%%%%%%%%%%%%%
Fig.\ref{fig:pt} shows the survival probability \eqref{survival} for the four cases considered above for times up to $6t_a$, with the same parameters used in Fig.\ref{fig:wp6}.
Cases $(i)$ and $(ii)$ were calculated analytically, while $(iii)$ and $(iv)$ by numerically integrating Eqs.\eqref{prop} and \eqref{survival}. An additional time scale to the
three described above now becomes relevant. For times $t\leq t_d$ the Gaussian wave packets have not spread outside $I$ and so there is not decay and $P^{(1)}(t\leq t_d)=1$. Hence
we have that $t_d= (1-q)/3\sigma$. Note that $t_d$ can be smaller, larger or equal to $t_w$, depending on the value of $q$. From Fig.\ref{fig:pt} we notice that $P^{(1)}(t)$ is
decreasing non-monotonically and shows strong oscillations. These are due to the spreading of the interference oscillations observed in Fig.\ref{fig:wp6}. Moreover, there are a
finite number of oscillations as for $t\gg1$, the frequency in the cosine term of Eq.\eqref{density} goes to zero. We can estimate the number of oscillations by $\int_{t_w}^{t_a}
\frac{q}{\pi t^2}d t= \frac{3}{\pi \sigma}-1$ for $\sigma\ll 1$.

For times $t\gg \sigma^2$, the two exponentials and the hyperbolic cosine of Eq.\eqref{density} are approximately equal to $1$. Using this approximation and integrating over $x\in
I$ we obtain an approximation for $P^{(1)}(t)$ for cases $(i)$ and $(ii)$
\begin{equation}
P^{(1)}(t)\approx \frac{2\sigma}{\sqrt{\pi}t} \prr{1 \mp \frac{\sigma}{2q}\sin\pr{ \frac{2q}{t} }}.
\end{equation}
Expanding for large times produces the expected asymptotics described by Eq.\eqref{NDasy}. Note that for case $(i)$ corresponding to $\mathcal{D}$ BC, the order $t^{-1}$ term cancels exactly,
so $P^{(1)}(t)\sim t^{-3}$.

\subsection{Asymptotic analysis}
In the limits $\eta\rightarrow0$ and $\eta\rightarrow\infty$ we obtain the propagators with $\mathcal{N}$ and $\mathcal{D}$ BCs respectively, as expected, with next-to leading order corrections:
\es{
K_{\eta}(x,x',t)&= K_{0}(x,x',t)\\
&-\textrm{erfc}\prr{\frac{\sqrt{-i}(x+x')}{\sqrt{2t}}}\eta+ \mathcal{O}(\eta^{2}), \label{asyeta1}
}
and
\es{
K_{\eta}(x,x',t)&=  K_{\infty}(x,x',t)\\
&- \frac{2 i (x+x')}{\eta t}K_{\textrm{free}}(x,-x',t)  + \mathcal{O}(\eta^{-2}). \label{asyeta2}
}
In the limit of $\eta\rightarrow-\infty$ however we obtain
\es{
K_{\eta}(x,x',t)&= -2|\eta| e^{-|\eta|(x+x')+\frac{1}{2}i t \eta^{2}} \\
 &+ K_{\infty}(x,x',t) + \mathcal{O}(\eta^{-1}), \label{asyeta3}
}
which implies that for negative (but finite) values of $\eta$, the single particle survival probability saturates at some constant, rather than decaying to zero, as observed in
Fig.\ref{fig:pt}. In the limit this constant vanishes and $K_{\infty}=K_{-\infty}$.

To understand this better we expand Eq.\eqref{Keta} for large times $t$ to get
\es{
K_{\eta}(x,x',t)&= (|\eta|-\eta) e^{\eta(x+x')+\frac{1}{2}i t \eta^{2}} \\
&-\frac{(1+i)(1+x \eta)(1+x' \eta)}{\sqrt{\pi} \eta^{2} t^{3/2}} + \mathcal{O}(t^{-5/2}). \label{asyt1}
}
Note that the first term vanishes when $\eta\geq0$. Eq.\eqref{asyt1} confirms that for $\eta>0$ the asymptotic single particle survival probability decays like $\sim t^{-3}$ (as
for $\mathcal{D}$ BC) and for $\eta<0$ the decay saturates at a constant $\mathcal{C}$. We concentrate on the first term in Eq.\eqref{asyt1} and apply it to a Gaussian initial state as in
\eqref{gaussian} to get that
\begin{equation}
\mathcal{C}= 4|\eta|\sigma e^{-2 q |\eta|}\pr{1- e^{-2|\eta|} }+\mathcal{O}(\sigma^2).
\end{equation}
For $\eta=-2$ and $\sigma=0.1$ we have that $\mathcal{C}\approx 0.126$ which is in very good agreement with the asymptotic decay of case $(iv)$ in Figure \ref{fig:pt}.

We remark that Eq.\eqref{asyt1} can be seemingly misleading, since for $\eta=0$ (corresponding to $\mathcal{N}$ BC) the second term diverges like $\sim -\sqrt{2i}/(\sqrt{\pi}\eta^{2}t^{3/2}
)$. Indeed this implies that the limit $\eta\rightarrow 0$ does not commute with that of $t \rightarrow \infty$. In fact we notice that Eq.\eqref{asyt1} reproduces the correct
asymptotics prescribed by Eq.\eqref{KN}, only if $\eta$ scales like $\sqrt{1/t}$. This can be understood in the following way: for $\eta<0$, the bound state $\chi$ becomes
completely delocalized in the limit $\eta\rightarrow0^{-}$. One can then imagine that $\chi$ acts like an ``$\eta$-thin carpet'' hindering the particle diffusion, thus giving a
slower decay of $P^{(1)}\sim t^{-1}$.

In brief, the exponent $\alpha$ of the asymptotic power law decay $P^{(1)}\sim t^{-\alpha}$ experiences discontinuous jumps from $0$ for $\eta<0$, to $1$ for $\eta=0$, and $3$ for
$\eta>0$. This unexpected discontinuity has interesting repercussions when considering the escape of $N\geq2$ particles from the interval $I=(0,1)$. These are discussed in the next
section.

\section{Many-particle case}\label{sec:many}
In the case of two particles, the initial state is given by Eq.\eqref{psisym} with
\es{
\psi_1(x)&= \pr{\frac{1}{\pi \sigma_{1}^{2}}}^{\frac{1}{4}} e^{-\frac{(x-q_1)^{2}}{2 \sigma_{1}^{2}}},\\
\psi_2(x)&= \pr{\frac{1}{\pi \sigma_{2}^{2}}}^{\frac{1}{4}} e^{-\frac{(x-q_2)^{2}}{2 \sigma_{2}^{2}}}
\label{gaussian2}
,}
and the wave function $\Psi^{(\bsn / \frm)}\left(x_1,x_2,t\right)$ at a later time $t>0$ is determined by Eqs.\eqref{K2} and \eqref{Psi2}. We have previously discussed the
asymptotic behavior of $P^{(2)}$ for $\mathcal{N}$ and $\mathcal{D}$ BCs (see Eq.\eqref{eq18}) and now turn to the general case of arbitrary $\eta$. This is achieved by expanding the propagator \eqref{K2}
in the limit of large $t$:
\es{
  \mathrm{Bosons, \;\;}\eta<0 :\;\; &K_\eta^{(\bsn)} \sim \textrm{const} \;\; \Rightarrow \; P^{(2)} \sim \textrm{const} \,, \\
  \mathrm{Fermions, \;\; }\eta<0 : \;\; &K_\eta^{(\frm)} \sim t^{-3/2} \;\; \Rightarrow \; P^{(2)} \sim t^{-3} \,,  \\
  \mathrm{Bosons, \;\;}\eta>0 : \;\; &K_\eta^{(\bsn)} \sim t^{-3} \;\; \Rightarrow \; P^{(2)} \sim t^{-6} \,,  \\ \mathrm{Fermions,\;\;}\eta>0 : \;\; &K_\eta^{(\frm)} \sim t^{-5}
\;\; \Rightarrow \; P^{(2)} \sim  t^{-10} \,.
\label{eq18b}}
It is interesting to note that the anti-symmetry of the fermionic wave function does not allow for a bound state and thus $P^{(2)}$ decays to zero. This completes the picture for
the two particle case.
%%%%%%%%%%%%%%%%%%%%%%%%%%%%%%%%%%%%%%%%%%%%%%%%%%%%%%%%%%%%%%%%%%%%%%%%%%%%%%%%%%%%%%%%%%%%%%%%%%%%%%%55
\begin{table}[h!]
{\small
\begin{center}
\caption{The asymptotic survival probability for one, two, and $N$ particles, as well as different BCs.}
\begin{tabular}{| c | c | c | c | c | c |}
\hline
 $\mathcal{R}$ BC  & $1$ &  $2$ bos  & $2$ fer & $N$ bos & $N$ fer\\ \hline
 $\eta<0$  & const & const & $ t^{-3}$ & const & $ t^{(1-N)(2N-1)}$ \\ \hline
 $\eta=0$  & $ t^{-1}$  & $ t^{-2}$ & $ t^{-6}$ & $ t^{-N}$ & $ t^{-N(2N-1)}$ \\ \hline
 $\eta>0$  & $ t^{-3}$  & $ t^{-6}$ & $ t^{-10}$ & $ t^{-3N}$ & $ t^{-N(2N+1)}$ \\ \hline
 $\eta=\pm\infty$ & $ t^{-3}$ & $ t^{-6}$& $ t^{-10}$ & $ t^{-3N}$ & $ t^{-N(2N+1)}$ \\ \hline
\end{tabular}
\label{tab}
\end{center}
}
\end{table}
%%%%%%%%%%%%%%%%%%%%%%%%%%%%%%%%%%%%%%%%%%%%%%%%%%%%%%%%%%%%%%%%%%%%%%%%%%%%%%%%%%%%%%%%%%%%%%%%%%%%%%%%%

The propagator \eqref{K2} can be generalized for $N\geq 2$ particles in a standard way \cite{M99}:
\es{
&K_{\eta}^{(\bsn / \frm)}(x_1,\ldots,x_N, x'_1,\ldots,x'_N,t)=\\
&\frac{1}{\sqrt{N!}}\sum_{i_1,\dots,i_N=1}^{N}\! \varepsilon_{i_1,\dots,i_N} K_{\eta}(x_1,x'_{i_1},t)\ldots K_{\eta}(x_N,x'_{i_N},t )
\label{det}
,}
where $\varepsilon_{i_1,\dots,i_N}$ is the totally antisymmetric (permutation) symbol in the case of fermions and is identically $1$ for bosons.

As done before, the asymptotic decay is obtained by expanding Eq.\eqref{det} for large times $t$. We present the results for $N=1,2,$ and for the general case in Table \ref{tab}
allowing for a direct comparison. In particular, we observe that the case of $\eta>0$ coincides with that of $\eta=\pm \infty$ corresponding to $\mathcal{D}$ BC. For the case of $\eta<0$, the
probability decays to a constant for bosons due to the presence of a bound state. Significantly, we observe that the case of $\eta=0$, corresponding to $\mathcal{N}$ BC, is special as it
separates the other cases in a discontinuous fashion for both bosons and fermions. We also observe that fermions always escape faster than bosons. As we have previously pointed
out, the asymptotic decay for two fermions with $\mathcal{N}$ BC is the same as for two bosons with $\mathcal{D}$ BC. However, this correspondence no longer holds for $N>2$. Finally, we note that the
survival probability of $N$ fermions with $\eta>0$ (or $\eta=\pm \infty$) has the same decay exponent as that of $N-1$ fermions with $\eta<0$.

\section{Conclusions}
We have addressed the influence of boundary conditions on the escape of $N$ indistinguishable particles in a one-dimensional setting, and have shown that they are equally important
as the prescribed quantum statistics. To this end, we have derived an exact closed form expression for the single-particle propagator on the positive real line in the presence of
Robin boundary conditions with a single control parameter $\eta$. This expression generalized existing results restricted to Dirichlet boundary condition ($\eta=\infty$) and
unveiled new non-trivial scenarios where escape may be completely suppressed. Moreover, we have found that the exponent of the asymptotic power law decay of the survival
probability is a discontinuous function of $\eta$. Our results hold for an arbitrary number of particles and are summarized in Table \ref{tab}.

In the light of recent atom-optics experiments addressing the dynamics of a small number of bosons and fermions in one dimension \cite{ZSLWRBJ12}, our findings may lead to new
applications in the area of quantum control and state manipulation.
On the theoretical side, it is also interesting to explore how higher dimensional waveguide-like \cite{JMT12} or network geometries \cite{CDG12},
external potentials, particle-particle interactions may affect the dynamics and in particular quantum escape.

\acknowledgments
The authors thank Chris Joyner and Susumu Shinohara for helpful discussions.
GG acknowledges support from the Ministry of Science, Serbia (Project III 45010).

\end{document}